\documentclass[sort&compress
    ,final            
  ]
  {aipproc}

\layoutstyle{8x11single}

\usepackage{amsfonts}
\usepackage{eucal}
\usepackage{bm}
\usepackage{amssymb}
\usepackage{amsmath}

\begin{document}

\title{Nonlocal Special Relativity: Amplitude Shift in Spin-Rotation Coupling}

\classification{03.30.+p, 04.20.Cv, 04.50.Kd, 11.10.Lm}
\keywords      {relativity, accelerated observers, nonlocality}

\author{Bahram Mashhoon}{address={Department of Physics and Astronomy, University of Missouri, Columbia, MO 65211, USA}
}



\begin{abstract}

The standard theory of relativity is based on the hypothesis of locality. The locality principle assumes that an object is affected only by its immediate surroundings and not by variables in the past. It follows that in standard relativity theory even wave properties are measured instantaneously. This contradicts the Bohr-Rosenfeld principle, according to which fields cannot be determined instantaneously. Nonlocal special relativity resolves the problem by taking past history into account. The current status of nonlocal electrodynamics is discussed and a new consequence of nonlocality, namely, a certain additional amplitude shift due to nonlocality in the spin-rotation coupling is presented.  
 
\end{abstract}

\maketitle


\section{Introduction}

The fundamental quantum laws of microphysics have all been formulated with respect to ideal inertial observers.  Such observers do not exist in nature; indeed, all realistic observers are more or less accelerated. Nevertheless, the basic significance of the hypothetical inertial observers for theoretical physics must be noted. These are the fundamental observers---namely, the free observers that remain at rest and thus characterize the global inertial frames in Minkowski spacetime~\cite{Ein, Min}. Their importance in the historical development of physics has been emphasized by I. B. Cohen~\cite{Coh}:

  ``I believe it fair to say that it was the freedom to consider problems either in a purely mathematical way or in a ``philosophical'' (or physical) way that enabled Newton to express the first law and to develop a complete inertial physics. After all, physics as a science may be developed in a mathematical way but it always must rest on experience---and experience never shows us pure inertial motion. Even in the limited examples of linear inertia discussed by Galileo, there was always some air friction and the motion ceased almost at once, as when a projectile strikes the ground. In the whole range of physics explored by Galileo there is no example of a physical object that has even a component of pure inertial motion for more than a very short time. It was perhaps for this reason that Galileo never framed a general law of inertia. He was too much a physicist.''

To extend the laws of physics to accelerated observers, a physical connection must be set up between accelerated and ideal inertial observers in Minkowski spacetime. The standard theory of relativity is based on the following pointwise relation: the accelerated observer is equivalent---at each event along its world line---to an otherwise identical momentarily comoving inertial observer. However, such a \emph{hypothesis of locality} is a relic of Newtonian mechanics of point particles and must be generalized for fields to a \emph{nonlocal} ansatz, which then leads naturally to a nonlocal special relativity theory~\cite{Mas}. From this acceleration-induced nonlocality, one would then expect a nonlocal general relativity theory as well, since inertia and gravitation are intimately linked in accordance with the principle of equivalence of inertial and gravitational masses. Recently, it has been possible to develop a nonlocal generalization of Einstein's theory of gravitation via the teleparallel equivalent of general relativity. The resulting nonlocal theory implies that gravity is nonlocal even in the Newtonian regime. The nonlocally modified Newtonian gravitation appears to provide a natural explanation for the dark matter problem in cosmology; in fact, \emph{nonlocal gravity simulates dark matter}. Further consequences of this theory are under active investigation at the present time~\cite{He1, He2, Blo, Bah, Car}. It is important to emphasize that in the nonlocal theory, the fields are always local; however, they satisfy nonlocal integro-differential equations. 

Nonlocal aspects of the gravitational interaction have also been explored by Novello and his collaborators~\cite{No1, Nov, YNO, Luc, Duq}. Novello's approach has involved an antisymmetric tensor of the third rank that was first introduced by Fierz in 1939~\cite{No1, Nov}. The spin-two field theory based on the Fierz tensor has its origin in the teleparallel approach to general relativity~\cite{YNO}. In this way, Novello \emph{et al.} have developed a parallel framework for a nonlocal theory of gravity~\cite{Duq}. \emph{It is a great pleasure for me to dedicate this paper to M\'ario Novello on the occasion of his seventieth birthday}. 

The plan of this paper is as follows. The essential steps in the argument for nonlocality are reviewed in the following section. The difficulties associated with nonlocal electrodynamics are then described. Observational results will be essential in the determination of the appropriate nonlocal kernel in this case.  As an example, a specific consequence of nonlocality for \emph{electromagnetic fields}, namely, the contribution of nonlocality to the amplitude shift in the spin-rotation coupling is treated in detail. The final section contains a brief discussion of our results. 

\section{Bohr-Rosenfeld Principle and Nonlocality}

Consider Maxwell's electrodynamics in an inertial frame of reference in Minkowski spacetime. The electric and magnetic fields, $\mathbf{E}(t,\mathbf{x})$ and $\mathbf{B}(t,\mathbf{x})$, respectively, that satisfy Maxwell's equations are assumed to be fields measured by the fundamental observers at rest in the inertial frame. In 1933, Bohr and Rosenfeld pointed out that in fact only \emph{spacetime averages} of these fields have immediate physical significance. That is, $\mathbf{E}(t,\mathbf{x})$ and $\mathbf{B}(t,\mathbf{x})$ occur in Maxwell's equations as idealizations~\cite{Boh, Ros}. To illustrate this point, Bohr and Rosenfeld considered a simple situation involving the measurement of the electric field using a macrophysical object  of volume $V$ and typical spatial dimension $L$, $V \sim L^3$, with \emph{uniform} volume charge density $\rho$. When placed in the external electric field  $\mathbf{E}(t,\mathbf{x})$, the object moves with respect to the fundamental observers according to the Lorentz force law, namely,       
\begin{equation}\label{1}
\frac{d\mathbf{P}}{dt}=\rho \int_V\mathbf{E}(t,\mathbf{x})~d^3x\,.
\end{equation}
Suppose that the motion of the object is monitored over an interval of time $T=t'' - t'$ and the momentum $\mathbf{P}$ is measured to be  $\mathbf{P'}$ and  $\mathbf{P''}$ at the initial and final instants, $t'$ and $t''$, respectively, of the experiment. Then, 
\begin{equation}\label{2}
\mathbf{P''}-\mathbf{P'}=\rho \int_{t'}^{t''}\int_V\mathbf{E}(t,\mathbf{x})~d^3x~dt=\rho \langle \mathbf{E} \rangle TV \,,
\end{equation}
where the measured electric field is
\begin{equation}\label{3}
 \langle \mathbf{E} \rangle=\frac{1}{\Delta} \int_{\Delta} \mathbf{E}(x)~ d^4x
\end{equation}
with $\Delta=TV$ and $x^{\mu}= (ct, \mathbf{x})$.  Henceforth we use units such that $c=1$, unless specified otherwise; moreover, the signature of the spacetime metric is $+2$ in our convention. It is assumed here that the times needed by the fundamental observers for momentum measurements are $\ll T$ and the corresponding displacements caused by these measurements are $\ll L$.

\emph{The gist of the Bohr-Rosenfeld argument is that fields cannot be measured instantaneously.} While this argument appears to be relatively innocuous for classical field measurements via ideal inertial observers, it leads to nonlocal special relativity for accelerated observers in Minkowski spacetime as a direct consequence of the existence of invariant acceleration scales.  We recall that an accelerated observer is generally endowed with intrinsic acceleration lengths such as $c^2/g(\tau)$ and $c/\Omega(\tau)$, where $\tau$ is the proper time of the observer and $g$ is the  magnitude of its translational acceleration, while $\Omega$ is the angular speed of rotation of its spatial frame with respect to a nonrotating (i.e., Fermi-Walker transported) frame. Let $\lambda^{\mu}{}_{(\alpha)}$ be the observer's orthornormal tetrad frame along its world line, then in general
\begin{equation}\label{4}
\frac{d\lambda^{\mu}{}_{(\alpha)}}{d\tau}=\Phi_{(\alpha)}{}^{(\beta)}(\tau)~\lambda^{\mu}{}_{(\beta)}\,,
\end{equation}
where $\Phi_{(\alpha)(\beta)} = - \Phi_{(\beta)(\alpha)}$ is the spacetime-invariant antisymmetric acceleration tensor of the observer. Here, in analogy with electrodynamics,  the translational and rotational accelerations of the observer constitute the spacetime-invariant ``electric'' and ``magnetic'' components of the acceleration tensor $\Phi_{(\alpha)(\beta)}$.

Let $\psi(x)$  be a basic radiation field as determined by the fundamental inertial observers and let $\Psi(x)$ be the field measured by the accelerated observer. According to the locality principle of the standard relativity theory, 
\begin{equation}\label{5}
 \Psi (\tau)=  \hat{\psi}(\tau) , \qquad  \hat{\psi} = \Lambda \psi\,,
\end{equation}
where $\hat{\psi}$ is the field measured instantaneously by the infinite set of hypothetical momentarily comoving inertial observers whose straight world lines are tangent to the world line of the accelerated observer. Moreover, at each instant $\tau$, $\hat{\psi}$ is related to $\psi$ by an element $\Lambda$ of a matrix representation of the Lorentz group.  To satisfy the Bohr-Rosenfeld principle, however, we must generalize Eq.~\eqref{5}. The most general \emph{linear} relationship between $\Psi$   and $\hat{\psi}$ that is consistent with causality is 
\begin{equation}\label{6}
 \Psi(\tau)=  \hat{\psi}(\tau) +  u(\tau - \tau_{0})\int_{ \tau_{0}}^{\tau}K(\tau,\tau') \hat{\psi}(\tau')d\tau' \,.
\end{equation}
Here, $u(t)$ is the unit step function such that $u(t) = 0$ for $t<0$ and $u(t) = 1$ for $t>0$ and $\tau_{0}$ is the initial instant of proper time at which the observer is accelerated.  As in the case of the Bohr-Rosenfeld Eq.~\eqref{3}, Eq.~\eqref{6} is manifestly covariant under the inhomogeneous Lorentz group of spacetime transformations.  It remains to determine kernel $K$, which must be proportional to the acceleration of the observer. This is the main problem of nonlocal special relativity. 

The basic nonlocal ansatz~\eqref{6} is a Volterra integral equation of the second kind. Therefore, it has the fundamental property that the relationship between $\Psi(\tau)$ and $\psi(\tau)$ is unique in the space of functions of physical interest in accordance with the Volterra-Tricomi theorem~\cite{Vol, Tri}. For a \emph{pure radiation field}, we impose the requirement that a \emph{constant} $\Psi$ should uniquely correspond to a \emph{constant} $\psi$; then, a variable $\psi$  will always lead to a variable $\Psi$. In this way, \emph{no observer can ever stay completely at rest with a pure radiation field}. Following this line of thought, we finally arrive at the kernel~\cite{BaM, BMa, Chi, ChM, Hel}
\begin{equation}\label{7}
K(\tau,\tau')=k(\tau')=-\frac{d\Lambda(\tau')}{d\tau'}\Lambda^{-1}(\tau')\,.
\end{equation}            
With this kernel, our nonlocal ansatz~\eqref{6} takes the form
\begin{equation}\label{8}
\Psi(\tau)=  \hat{\psi}(\tau_{0}) + \int_{ \tau_{0}}^{\tau}\Lambda(\tau')\frac{d\psi(\tau')}{d\tau'}d\tau' \,
\end{equation}
for $\tau \ge \tau_{0}$. It turns out that the nonlocal contribution in our ansatz~\eqref{6} is negligible when the intrinsic scale of the phenomenon under observation is sufficiently small in comparison with the scale of variation of the state of the observer. This happens to be the case for most Earth-based experiments, since $c^{2}/|\mathbf{g}_{\oplus}|\approx 1$ light year and $c/|\boldsymbol{\Omega}_{\oplus}|\approx 28$ astronomical units.  

The implications of the nonlocal theory have been worked out in detail---see, for instance, Ref.~\cite{BMA} for the nonlocal Dirac equation. Once acceleration is turned off, its memory in general persists in the form of an additive constant field. In nonlocal special relativity, local fields satisfy integro-differential field equations that contain the memory of past acceleration. For fundamental scalar or pseudoscalar radiation fields, $\Lambda=1$ and hence $K=0$; indeed, nonlocal special relativity predicts that such radiation fields do not exist in nature. This prediction is consistent with observation. 

The nonlocal kernel given by Eq.~\eqref{7} has been derived for a \emph{pure radiation field}. Its application to the electromagnetic field encounters difficulties, however, as it is not clear how one must treat nonlocal electrostatics and magnetostatics in this framework. This problem is the subject of the next section. 

\section{Nonlocal Electrodynamics}

Consider an electromagnetic field $F_{\mu\nu}(x)$  and the corresponding gauge potential $A_{\mu}(x)$, $F_{\mu\nu}=\partial_{\mu}A_{\nu}-\partial_{\nu}A_{\mu}$, in the global inertial frame in Minkowski spacetime.  Along the world line of the accelerated observer, the fields measured by the momentarily comoving inertial observers are
\begin{equation}\label{9}
A_{(\alpha)}(\tau)=A_{\mu}\lambda^{\mu}{}_{(\alpha)}, \qquad  F_{(\alpha)(\beta)}(\tau)=F_{\mu\nu}\lambda^{\mu}{}_{(\alpha)}\lambda^{\nu}{}_{(\beta)}\,.
\end{equation}
Let $\mathcal{A}_{(\alpha)}(\tau)$ and $\mathcal{F}_{(\alpha)(\beta)}(\tau)$ be the fields determined by the accelerated observer; then, in accordance with our nonlocal ansatz, 
\begin{eqnarray}\label{10}
\mathcal{A}_{(\alpha)}(\tau)=A_{(\alpha)}(\tau)+u(\tau-\tau_{0})\int_{\tau_{0}}^{\tau}K_{(\alpha)}{}^{(\beta)}(\tau,\tau')A_{(\beta)}(\tau')d\tau'\,
\end{eqnarray}
and
\begin{eqnarray}\label{11}
\mathcal{F}_{(\alpha)(\beta)}(\tau)=F_{(\alpha)(\beta)}(\tau)+u(\tau-\tau_{0})\int_{\tau_{0}}^{\tau}K_{(\alpha)(\beta)}{}^{(\gamma)(\delta)}(\tau,\tau')
F_{(\gamma)(\delta)}(\tau')d\tau'\,.
\end{eqnarray}
If the kernel in Eq.~\eqref{10} is chosen in accordance with Eq.~\eqref{7}, namely, 
\begin{eqnarray}\label{12}
K_{(\alpha)}{}^{(\beta)}(\tau,\tau')=-~\Phi_{(\alpha)}{}^{(\beta)}(\tau')\,,
\end{eqnarray}
then a constant $A_{\mu}(x)$ will be determined to be constant by all accelerated observers. This circumstance poses no difficulty, as the electromagnetic field vanishes altogether in this case. However, the situation is quite different if the kernel in Eq.~\eqref{11} is chosen in accordance with Eq.~\eqref{7}; then, constant electromagnetic fields in the laboratory will always be measured to be constant by any accelerated observer. This conclusion appears to contradict the results of Kennard's experiment~\cite{Ken, Peg}. The issue has been discussed in detail in~\cite{MaN}; in this case, the kernel must be determined from $\Phi_{(\alpha)(\beta)}$, the Minkowski metric tensor and the Levi-Civita tensor (with $\epsilon_{0123}=1$ in our convention). A simple choice for the kernel in Eq.~\eqref{11} turns out to be a linear combination of the kernel given by Eq.~\eqref{7}, namely,
\begin{equation}\label{13}
\kappa_{(\alpha)(\beta)}{}^{(\gamma)(\delta)}=-2\Phi_{\lbrack(\alpha)}{}^{\lbrack(\gamma)}~\delta_{(\beta)\rbrack}{}^{(\delta)\rbrack}
\end{equation}
and its dual
\begin{equation}\label{14}
\kappa^*_{(\alpha)(\beta)}{}^{(\gamma)(\delta)}=\frac{1}{2}\epsilon_{(\alpha)(\beta)}{}^{(\rho)(\sigma)}~\kappa_{(\rho)(\sigma)}{}^{(\gamma)(\delta)}\,.
\end{equation}
That is, we tentatively assume that in vacuum
\begin{equation}\label{15}
K_{(\alpha)(\beta)}{}^{(\gamma)(\delta)}(\tau,\tau^{'})= p~\kappa_{(\alpha)(\beta)}{}^{(\gamma)(\delta)}(\tau')+ q~\kappa^*_{(\alpha)(\beta)}{}^{(\gamma)(\delta)}(\tau')\,,
\end{equation}
where $p$ and $q$ are real constant coefficients to be determined from experiment~\cite{MaN} . 

Some of the properties of kernel~\eqref{13} have been discussed in~\cite{MaN, Mas}. In particular, the right dual of kernel~\eqref{13} is equal to its left dual, while its mixed duals vanish; that is, kernel~\eqref{13} has a \emph{unique} dual and this property is a direct consequence of the relation $\Phi_{(\alpha)(\beta)}=-\Phi_{(\beta)(\alpha)}$. It is interesting to define
\begin{equation}\label{15a}
\chi_{(\alpha)(\beta)}{}^{(\gamma)(\delta)}=-2\Phi^*_{\lbrack(\alpha)}{}^{\lbrack(\gamma)}~\delta_{(\beta)\rbrack}{}^{(\delta)\rbrack},
\end{equation}
where $\Phi^*_{(\alpha)(\beta)}$ is the dual of the acceleration tensor,
\begin{equation}\label{15b}
\Phi^*_{(\alpha)(\beta)}=\frac{1}{2}\epsilon_{(\alpha)(\beta)}{}^{(\gamma)(\delta)}~\Phi_{(\gamma)(\delta)}\,.
\end{equation}
This tensor is antisymmetric; therefore, kernel~\eqref{15a} has a unique dual as well, and it is then straightforward to show, using a generalized Kronecker delta~\cite{Hel}, that
\begin{equation}\label{15c}
\chi^*_{(\alpha)(\beta)}{}^{(\gamma)(\delta)}=-\kappa_{(\alpha)(\beta)}{}^{(\gamma)(\delta)}\,.
\end{equation}
Taking the dual of this equation, we find
\begin{equation}\label{15d}
\chi_{(\alpha)(\beta)}{}^{(\gamma)(\delta)}=\kappa^*_{(\alpha)(\beta)}{}^{(\gamma)(\delta)}\,.
\end{equation}
Inspection of Eqs.~\eqref{15c} and~\eqref{15d} reveals that under the double duality operation,  kernels $\kappa_{(\alpha)(\beta)}{}^{(\gamma)(\delta)}$ and $\kappa^*_{(\alpha)(\beta)}{}^{(\gamma)(\delta)}$---and hence the nonlocal electromagnetic kernel~\eqref{15}---are simply multiplied by $-1$; for instance, 
\begin{equation}\label{15e}
\frac{1}{2}\epsilon_{(\alpha)(\beta)}{}^{(\rho)(\sigma)}~\kappa^*_{(\rho)(\sigma)}{}^{(\gamma)(\delta)}=-\kappa_{(\alpha)(\beta)}{}^{(\gamma)(\delta)}\,.
\end{equation}

Available experimental results regarding electrodynamics of accelerated media are rather meager and possibly unreliable; in any case, the effect of nonlocality is expected to be very small and hence rather difficult to detect. One may therefore try to find indirect evidence for nonlocality. For instance, in the correspondence limit of large quantum numbers, electrons in atoms can be regarded as following on average classical accelerated paths and one can study their behavior under the influence of incident electromagnetic radiation. In particular, it turns out that the impulse approximation of quantum theory corresponds to the hypothesis of locality. A detailed investigation reveals that in the correspondence limit, quantum results are in better qualitative agreement with the predictions of the nonlocal theory than with the standard relativity theory based on the locality principle~\cite{BMN}.

Some of the observational consequences of kernel~\eqref{15} have been worked out in~\cite{MaN, Mas}. It is argued there that we should assume $p\ge0$ and $q\ne0$; moreover, we expect that $|q|\ll 1$, since $q$ is associated with violations of parity and time reversal invariance. A further implication of this kernel involves the amplitude shift in spin-rotation coupling, to which we now turn. 

\section{Spin-Rotation Coupling and Nonlocality}

The nonlocal aspects of spin-rotation coupling in electrodynamics have been the subject of recent studies~\cite{BM1, BM2}. On the other hand, starting from classical electrodynamics and kernel~\eqref{15}, we are interested here in the measurement of the electromagnetic field when a circularly polarized plane monochromatic wave of frequency $\omega$ is normally incident on an observer that rotates uniformly with frequency $\Omega > 0$ about the direction of incidence of the  radiation. Fourier analysis of 
$\mathcal{F}_{(\alpha)(\beta)}(\tau)$ implies in this case that the frequency measured by the rotating observer is $\gamma~(\omega\mp \Omega)$, where $\gamma$ is the Lorentz factor of the rotating observer and the upper (lower) sign refers to an incident positive (negative) helicity wave. The only exception occurs in the resonance case of incident positive helicity wave when $\omega=\Omega$, in which case the complex amplitude of the measured field varies with time as $1-i(p+iq)\gamma~\Omega\tau$; this feature is hence a direct consequence of nonlocality~\cite{MaN}.

The complex amplitude of the incident circularly polarized radiation, as measured by the rotating observer, also depends upon the helicity of the radiation as a direct consequence of nonlocality; indeed, this dependence away from resonance is given by~\cite{MaN}
\begin{eqnarray}\label{16}
1+(\pm p+iq)\frac{\Omega}{\omega\mp \Omega}\,.
\end{eqnarray}

The general phenomenon of spin-rotation coupling, which is due to the inertia of intrinsic spin, has helped elucidate the energy shift that is observed when a spinning particle passes through a \emph{rotating} spin flipper~\cite{Neu, Kai}. For electromagnetic radiation, the \emph{frequency shift} that occurs when circularly polarized radiation of frequency $\omega$ passes through a rotating device (with rotation frequency $\Omega \ll \omega$) that flips the helicity of the radiation has been known observationally for a long time and has been simply explained via the photon picture~\cite{All, GAr, Gar, Sim, Bag, Pip, Nie}.  Let $N$ photons pass through the nonrelativistic device and undergo helicity flip; hence, the magnitude of the angular momentum of the rotating device $\mathcal{J}$ must change by $\delta \mathcal{J}=2N\hbar$, as a consequence of the law of conservation of angular momentum. Since the device is rotating with frequency $\Omega$, its energy $\mathcal{E}$ must change by $\delta \mathcal{E} = \Omega ~ \delta \mathcal{J}$ or $2N\hbar \Omega$. It then follows from energy conservation that each photon must suffer an energy shift equal to $2 \hbar \Omega$ and hence a corresponding frequency shift equal to $2\Omega$. 

In connection with the phenomenon of frequency shift, it is important to remark that \emph{nonlocality} brings about a corresponding extra \emph{amplitude shift} as well. This is described in the next section by means of nonlocal classical electrodynamics using kernel~\eqref{15}.

\section{Amplitude Shift}

Imagine, for instance, a simple situation involving the passage of light wave of definite helicity propagating along the $z$ axis through a rotating half-wave plate (HWP) as in Figure 1. In the \emph{background global inertial frame}, the incident positive-helicity wave has initial frequency $\omega_{i}$ and constant amplitude $\alpha_{i}$, while the corresponding outgoing negative-helicity wave has final frequency $\omega_{f}$ and constant amplitude $\alpha_{f}$. We assume here that $\gamma \approx 1$, $\omega_i \gg \Omega$ and terms proportional to $(\Omega/\omega_i)^2$ are negligible, so that we can---among other things---approximate $\omega\mp \Omega$ in Eq.~\eqref{16} by $\omega$. At the \emph{rotating outer boundaries of the HWP}, imagine observers at rest in the rotating frame of the HWP; according to these observers, the measured frequencies and amplitudes are then given as in the previous section by (cf. Figure 1)

\begin{eqnarray}\label{17}
\omega'_{1}\approx\omega_{i}-\Omega\,, \qquad  \alpha'_{1}\approx \alpha_{i}~\Big \lbrack1+(p+iq)\frac{\Omega}{\omega_{i}}\Big \rbrack
\end{eqnarray}
and
\begin{eqnarray}\label{18}
\omega'_{2}\approx\omega_{f}+\Omega\,, \qquad  \alpha'_{2}\approx \alpha_{f}~\Big \lbrack1+(-p+iq)\frac{\Omega}{\omega_{f}}\Big \rbrack\,.
\end{eqnarray}
Here we have neglected time dilation; indeed, all terms beyond the linear order in $\Omega$ are neglected in this analysis.
   
\begin{figure}
\includegraphics[height= 0.5\textheight]{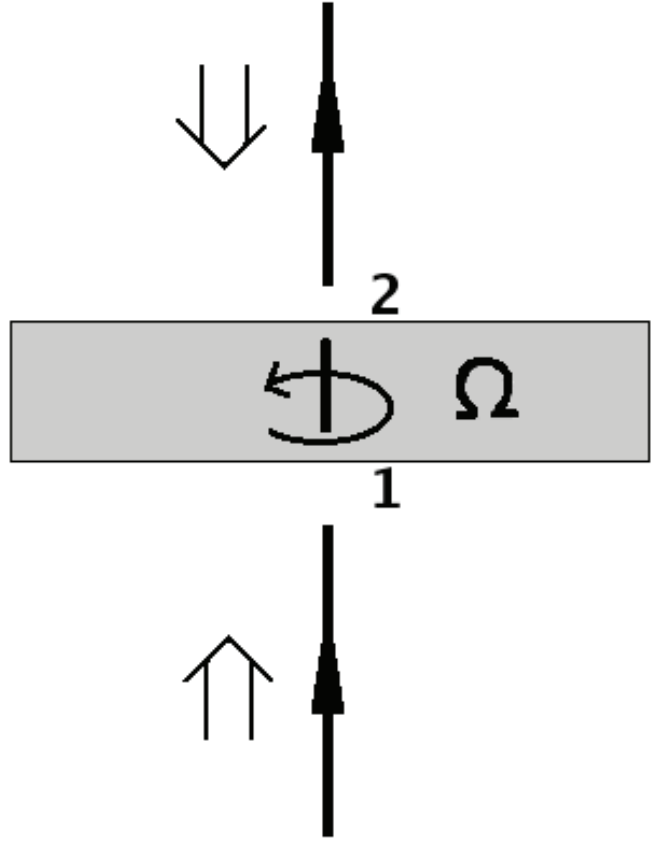}
\caption{Schematic diagram for the passage of incident positive-helicity radiation through a rotating HWP.}
\label{Figure 1}
\end{figure}

The rotation of the HWP is assumed to be uniform; therefore, the wave frequency remains constant in the rotating frame, namely, $\omega'_{1}=\omega'_{2}$. Moreover, $\alpha'_{2}=\mathcal{T}\alpha'_{1}$, where $\mathcal{T}$ is the net transmission amplitude through the uniformly rotating HWP as measured \emph{in the rotating frame}. Thus Eqs.~\eqref{17} and~\eqref{18} imply that 

\begin{eqnarray}\label{19}
\omega_{f}\approx \omega_{i}-2\Omega\,,
\end{eqnarray}
which is the expected \emph{frequency shift}, and
\begin{eqnarray}\label{20}
\alpha_{f}\Big \lbrack1+(-p+iq)\frac{\Omega}{\omega_{f}}\Big \rbrack \approx \mathcal{T}\alpha_{i}\Big \lbrack1+(p+iq)\frac{\Omega}{\omega_{i}}\Big \rbrack\,.
\end{eqnarray}

The response of the \emph{rotating} HWP to the passage of radiation could in general be complicated~\cite{Pip, Bia, Hau}. On the other hand, a static HWP can be simply treated in the standard classical manner~\cite{Bor}. For the calculation of $\mathcal{T}$, we adopt an elementary approach that should be adequate for a sufficiently narrow incident beam along the axis of rotation of a slowly rotating HWP, namely, in the rest frame of the HWP, the passage of radiation of frequency $\omega'=\omega'_{1}=\omega'_{2}$ through the HWP is treated just as in the absence of rotation~\cite{Bor}. In the horizontal $(x,y)$ plane of the HWP, the slow axis (with principal index of refraction $n_s$) is perpendicular to the fast axis (with principal index of refraction $n_f$) such that $\delta n= n_s-n_f>0$ is sufficiently small. The thickness $D$ of the HWP along the vertical $z$ direction is then connected to its birefringence $\delta n$ via the characteristic relation $\omega' \delta n~D/c= \pi +2\pi m$, where $m$ is a positive integer. 

In the background inertial frame, the circular polarization states of the radiation can be expressed in terms of unit vectors as $(\hat {\mathbf{x}}\pm i\hat{\mathbf{y}})/\sqrt{2}$~\cite{MaN}. In the rotating frame, which is the rest frame of the HWP, the corresponding unit vectors are $\approx (\hat {\mathbf{x'}}\pm i\hat{\mathbf{y'}})/\sqrt{2}$ for $\gamma \approx 1$~\cite{MaN}. Furthermore, we assume that  the rotating $x' (y')$ axis coincides with the slow (fast) axis. Using the formulas given in~\cite{Bor}, the transmission amplitudes for light that is \emph{linearly polarized} along the slow and fast axes can then be calculated and we find that $\mathcal{T}_s \approx -\mathcal{T}_f$; hence, for \emph{circularly polarized} radiation
\begin{eqnarray}\label{21}
\mathcal{T}\approx \mathcal{T}_s=\frac{e^{i\zeta_s}}{1-if(n_s)e^{i\zeta_s}\sin \zeta_s}\,,
\end{eqnarray}
where
\begin{eqnarray}\label{22}
\zeta_s=\frac{1}{c}\omega' n_s~D, \qquad f(n) = \frac{(n-1)^2}{2n}\,.
\end{eqnarray}
Here, the refractive index of the slow axis $n_s$ is assumed to be slightly larger than the refractive index of the fast axis $n_f$ in such a way that $f(n_s)\approx f(n_f)$. More precisely, calculating $\delta f=f(n_s)-f(n_f)$ using Taylor expansion and assuming that $\delta f\ll f$, we find that $\delta n$ must satisfy the condition
\begin{eqnarray}\label{23}
\frac{\delta n}{n_f}\ll\frac{n_f-1}{n_f+1}\,.
\end{eqnarray}
Moreover, we note that $\exp (i\zeta)\sin \zeta$ in the denominator of Eq.~\eqref{21} is invariant under $\zeta \mapsto \zeta \pm (\pi +2\pi m)$.

It follows from Eqs.~\eqref{19} and~\eqref{20} that the emerging amplitude is enhanced due to nonlocality, namely, $\alpha_{f} \approx \mathcal{T}~\Sigma~\alpha_{i}$, where
\begin{eqnarray}\label{24}
\Sigma(\Omega)\approx 1+p\frac{2\Omega}{\omega_{i}}\,.
\end{eqnarray}
That is, $q$ drops out at the linear order in $\Omega$ and $\Sigma(\Omega)-1$ is the relative \emph{amplitude shift due to nonlocality} given approximately by $2p~\Omega / \omega_{i}$.
Assuming that $p>0$ is, say, of the order of unity, the amplitude upshift is expected to be very small compared to unity. For an incident negative-helicity wave, $\Omega\mapsto -\Omega$ in Eqs.~\eqref{17} and~\eqref{18} and so there would be an upshift in frequency by $2\Omega$ and a corresponding downshift in amplitude, so that the emerging amplitude would be diminished due to nonlocality by the  factor $\Sigma(-\Omega)\approx 1-2p~\Omega/\omega_{i}$. The frequency shift given by Eq.~\eqref{19} is a general consequence of spin-rotation coupling; indeed, as mentioned in the previous section, such a shift is experimentally well known and has been the subject of a number of investigations---see, for instance,~\cite{All, GAr, Gar, Sim, Bag, Pip, Nie, Cou, Bas, Bl1, Bl2, Bl3} and the references cited therein. The situation is different, however, for the amplitude shift, which has not yet been detected. In fact, nonlocality implies that for $p>0$ the corresponding relative shift in amplitude is positive (negative) when the helicity of the incident wave is in the same (opposite) sense as the rotation of the HWP. 

The amplitude shift due to nonlocality occurs in addition to the amplitude shift that comes about as a direct result of spin-rotation coupling and is revealed through the dependence of the transmission amplitude $\mathcal{T}$ upon $\omega' \approx \omega_i -\Omega$ via $\zeta$; indeed, this dependence can be clearly seen in the transmission coefficient,
\begin{eqnarray}\label{25}
|\mathcal{T}|^2\approx \frac{1}{1+f(f+2)\sin^2 \zeta}\,,
\end{eqnarray}
where $f(f+2)=(n^2-1)^2/(4n^2)$, $\zeta \approx \omega_i~n~D/c-\Omega~n~D/c$ and $n$ can be either $n_s$ or $n_f$.

For visible light with $\omega_{i}/(2\pi)\approx 5\times 10^{14}$ Hz and a HWP rotating uniformly with frequency $\Omega/(2\pi)\approx 25$ Hz, we have $2\Omega/\omega_{i}\approx 10^{-13}$, which implies a rather small relative shift in amplitude, too small perhaps to be detectable at present. This is consistent with the relatively low level of amplitude sensitivity of current observational data; in fact, experiments of this type have not reported any similar amplitude shift---see~\cite{All, GAr, Gar, Sim, Bag, Pip, Nie, Cou, Bas, Bl1, Bl2, Bl3} and the references cited therein. Perhaps the amplitude shift due to nonlocality would be easier to detect with microwaves or radio waves.

\section{Discussion}

The main conceptual steps that lead to a nonlocal theory of relativity have been outlined in this paper. The problems associated with nonlocal electrodynamics have to do with the determination of the corresponding nonlocal kernel. In this regard, future experimental results concerning electrodynamics of accelerated systems will be decisive. 

A new consequence of nonlocal electrodynamics, namely, an extra amplitude shift proportional to $p$ in spin-rotation coupling is pointed out. Estimates suggest that this direct result of nonlocality is negligibly small in most situations. The possibility of detecting this novel effect is briefly discussed.


\begin{theacknowledgments}
This paper is based in part on lectures delivered at the XIV Brazilian School of Cosmology and Gravitation (August 30 - September 11, 2010). Thanks are due to M\'ario Novello and the organizing committee for their kind invitation and excellent hospitality. I am grateful to Friedrich Hehl for helpful discussions on all aspects of nonlocal electrodynamics.
\end{theacknowledgments}

\bibliographystyle{aipproc}



\bibliographystyle{aipproc}   

\bibliography{sample}

\begin{thebibliography}{99}


\bibitem{Ein} 
A.~Einstein, \textit{The Meaning of Relativity}, Princeton University Press, Princeton, NJ, 1955.

\bibitem{Min} 
H.~Minkowski, in \textit{The Principle of Relativity}, by H.~A.~Lorentz, A.~Einstein, H.~Minkowski and H.~Weyl, Dover, New York, 1952.

\bibitem{Coh}
I.~ B.~Cohen, \textit{The Birth of a New Physics}, Doubleday Anchor Books, Garden City, NY, 1960, pp. 164--165. 

\bibitem{Mas} 
B.~Mashhoon, 
\textit{Ann.\ Phys.\ (Berlin)} {\bf 17}, 705 
(2008) [arXiv:0805.2926 [gr-qc]].

\bibitem{He1}
 F.~W.~Hehl and B.~Mashhoon, \emph{Phys. Lett. B}  \textbf{673}, 279 (2009) [arXiv:0812.1059 [gr-qc]].
              
\bibitem{He2}
F.~W.~Hehl and B.~Mashhoon, \emph{Phys. Rev. D}  \textbf{79}, 064028 (2009) [arXiv:0902.0560 [gr-qc]].

\bibitem{Blo}
H.-J.~Blome, C.~Chicone, F.~W.~Hehl and B.~Mashhoon, \emph{Phys. Rev. D} \textbf{81}, 065020 (2010) [arXiv:1002.1425 [gr-qc]].

\bibitem{Bah}
B.~Mashhoon,  ``Nonlocal Gravity,''  in  
\emph{Cosmology and Gravitation, Proc. XIV Brazilian School of Cosmology and Gravitation (Rio de Janeiro, 2010)}, edited by M.~Novello and S.~E.~Perez Bergliaffa, Cambridge Scientific Publishers, UK, 2011, pp. 1--9 [arXiv:1101.3752 [gr-qc]].

\bibitem{Car}
C.~Chicone and B.~Mashhoon, \emph{J. Math. Phys.} {\bf 53}, 042501 (2012) [arXiv:1111.4702 [gr-qc]].

\bibitem{No1}
M.~Novello, arXiv: gr-qc/0212026.

\bibitem{Nov}
M.~Novello and R.~P.~Neves, \emph{Class. Quantum Grav.} {\bf 19}, 5335 (2002).

\bibitem{YNO}
Yu.~N.~Obukhov and J.~G.~Pereira,
\emph{Phys. Rev. D} {\bf 67}, 044008 (2003). 

\bibitem{Luc}
Luciane R.~de Freitas, M.~Novello and N.~Pinto-Neto, \emph{J. Math. Phys.} {\bf 35}, 734 (1994).

\bibitem{Duq}
M.~Novello and S.~L.~S.~Duque,
\emph{Int.\ J.\ Mod.\ Phys.\ D} {\bf 4}, 79 (1995).

\bibitem{Boh} 
N.~Bohr and L.~Rosenfeld, \emph{K. Dan. Vidensk. Selsk. Mat. Fys. Medd.}  \textbf{12}, No. 8 (1933);\\ translated in \textit{Quantum Theory and Measurement}, edited by J.~A.~Wheeler and W.~H.~Zurek, Princeton University Press, Princeton, NJ, 1983.
	
\bibitem{Ros} 
N.~Bohr and L.~Rosenfeld, \emph{Phys. Rev.}  \textbf{78}, 794 (1950).

\bibitem{Vol} 
V.~Volterra, \emph{Theory of Functionals and of Integral and Integro-Differential Equations}, Dover, New York, 1959.

\bibitem{Tri}
 F.~G.~Tricomi, {\it Integral Equations}, Interscience, New York, 1957.
 
 \bibitem{BaM}
 B.~Mashhoon, \emph{Phys. Rev. A}  \textbf{47}, 4498 (1993).
       
\bibitem{BMa}
B.~Mashhoon,  ``Nonlocal Electrodynamics,''  in \emph{Cosmology and Gravitation}, 
\emph{Proc. VII Brazilian School of Cosmology and Gravitation (Rio de Janeiro, August, 1993)},
edited by M.~Novello, Editions Fronti\`eres, Gif-sur-Yvette, 1994, pp. 245--295. 
\url{http://www.cbpf.br/~cosmogra/Escolas/ind_class_field.html}

\bibitem{Chi} 
C.~Chicone and B.~Mashhoon, \emph{Ann. Phys. (Berlin)}  \textbf{11}, 309 (2002).
        
\bibitem{ChM} 
C.~Chicone and B.~Mashhoon, \emph{Phys. Lett. A}  \textbf{298}, 229 (2002).

\bibitem{Hel}  
F.~W.~Hehl and Yu.~N.~Obukhov, {\it Foundations of
    Classical Electrodynamics: Charge, Flux, and Metric}, Birkh\"auser, Boston, MA, 2003.


\bibitem{BMA}
 B.~Mashhoon, \emph{Phys. Rev. A}  \textbf{75}, 042112 (2007) [arXiv: hep-th/0611319].
 
\bibitem{Ken}
E.~H.~Kennard,
{\it Phil. Mag.} {\bf 33}, 179 (1917).

\bibitem{Peg}
G.~B.~Pegram,
{\it Phys. Rev.} {\bf 10}, 591 (1917).

\bibitem{MaN}
 B.~Mashhoon, \emph{Phys. Lett. A}  \textbf{366}, 545 (2007) [arXiv: hep-th/0702074].
 
\bibitem{BMN}
B.~Mashhoon, \emph{Phys. Rev. A}  \textbf{72}, 052105 (2005) [arXiv: hep-th/0503205]. 

\bibitem{BM1}
B.~Mashhoon, \emph{Phys. Rev. A}  \textbf{79}, 062111 (2009) [arXiv:0903.1315 [gr-qc]].

\bibitem{BM2} 
B.~Mashhoon, \emph{Ann. Phys. (Berlin)} {\bf 523}, 226 (2011) [arXiv:1006.4150 [gr-qc]].

\bibitem{Neu}
B.~Mashhoon, R.~Neutze, M.~Hannam and G.~E.~Stedman, \emph{Phys. Lett. A} \textbf{249}, 161 (1998).
        
\bibitem{Kai}         
B.~Mashhoon and H.~Kaiser, \emph{Physica B} \textbf{385-386}, 1381 (2006) [arXiv: quant-ph/0508182].

\bibitem{All}
P.~J.~Allen, \emph{Am. J. Phys.} \textbf{34}, 1185 (1966).

\bibitem{GAr}
B.~A.~Garetz and S.~Arnold, \emph{Opt. Commun.} \textbf{31}, 1 (1979).

\bibitem{Gar}
B.~A.~Garetz, \emph{J. Opt. Soc. Am.} \textbf{71}, 609 (1981).

\bibitem{Sim}
R.~Simon, H.~J.~Kimble and E.~C.~G.~Sudarshan, \emph{Phys. Rev. Lett.} \textbf{61}, 19 (1988).

\bibitem{Bag}
V.~Bagini \textit{et al}., \emph{Eur. J. Phys.} \textbf{15}, 71 (1994).

\bibitem{Pip}
A.~B.~Pippard, \emph{Eur. J. Phys.} \textbf{15}, 79 (1994).

\bibitem{Nie}
G.~Nienhuis, \emph{Opt. Commun.} \textbf{132}, 8 (1996).

\bibitem{Cou}
J.~Courtial \textit{et al}., \emph{Phys. Rev. Lett.} \textbf{81}, 4828 (1998).

\bibitem{Bas}
I.~V.~Basistiy \textit{et al}., \emph{Opt. Lett.} \textbf{28}, 1185 (2003).

\bibitem{Bl1}
K.~Y.~Bliokh, Y.~Gorodetski, V.~Kleiner and E.~Hasman, \emph{Phys. Rev. Lett.} \textbf{101}, 030404 (2008).

\bibitem{Bl2}
K.~Y.~Bliokh \textit{et al}., \emph{Nature Photonics} \textbf{2}, 748 (2008).

\bibitem{Bl3}
K.~Y.~Bliokh, 
\emph{J. Opt. A: Pure Appl. Opt.} \textbf{11}, 094009 (2009).

\bibitem{Bia}
I.~Bialynicki-Birula and Z.~Bialynicka-Birula, \emph{Phys. Rev. Lett.} \textbf{78}, 2539 (1997).
    
\bibitem{Hau}
J.~C.~Hauck and B.~Mashhoon, \emph{Ann. Phys. (Berlin)} \textbf{12}, 275 (2003).

\bibitem{Bor}
M.~Born and E.~Wolf, \emph{Principles of Optics}, Pergamon Press, Oxford, 1975, pp. 61--66.

\clearpage
\end{thebibliography}

\IfFileExists{\jobname.bbl}{}
 {\typeout{}
  \typeout{******************************************}
  \typeout{** Please run "bibtex \jobname" to optain}
  \typeout{** the bibliography and then re-run LaTeX}
  \typeout{** twice to fix the references!}
  \typeout{******************************************}
  \typeout{}
 }

\end{document}